\documentstyle[aps, epsf, amsfonts]{revtex}

\begin{document}
\input epsf.tex

\normalsize  

\title{Computational Resources to Filter Gravitational Wave Data with P-Approximant Templates.}
\author{Edward K. Porter\footnote{Present address : Laboratoire de l'Acc\'el\'erateur Lin\'eaire, B.P. 34, B\^{a}timent 208, Campus d'Orsay, 91898 Orsay Cedex, France.  e-mail: porter@lal.in2p3.fr}}
\address{Dept. of Physics and Astronomy, Cardiff University, 5 The Parade, Cardiff, Wales, UK,
CF24 3YB}
\maketitle
\vspace{1cm}
\begin{abstract}
\noindent The prior knowledge of the gravitational waveform from compact binary systems makes matched filtering an attractive detection strategy.   This detection method involves the filtering of the detector output with a set of theoretical waveforms or templates.  One of the most important factors in this strategy is knowing how many templates are needed in order to reduce the loss of possible signals.  In this study we calculate the number of templates and computational power needed for a one-step search for gravitational waves from inspiralling binary systems.  We build on previous works by firstly expanding the post-Newtonian waveforms to 2.5-PN order and secondly, for the first time, calculating the number of templates needed when using P-approximant waveforms.  The analysis is carried out for the four main first-generation interferometers, LIGO, GEO600, VIRGO and TAMA.  As well as template number, we also calculate the computational cost of generating banks of templates for filtering GW data.  We carry out the calculations for two initial conditions. In the first case we assume a minimum individual mass of $1\,M_{\odot}$ and in the second, we assume a minimum individual mass of $5\,M_{\odot}$.  We find that, in general, we need more P-approximant templates to carry out a search than if we use standard PN templates.  This increase varies according to the order of PN-approximation, but can be as high as a factor of 3 and is explained by the smaller span of the P-approximant templates as we go to higher masses.  The promising outcome is that for 2-PN templates the increase is small and is outweighed by the known robustness of the 2-PN P-approximant templates.
\end{abstract}

\section{Introduction}
The inspiral phases of close binary systems are expected to be the most important sources of gravitational waves (GW) for the first generation of detectors, LIGO, GEO600, VIRGO and TAMA~\cite{LIGO,VIRGO,GEO,TAMA}.  Under radiation reaction the orbit of the binary decays emitting a signal which increases in amplitude and frequency called a ``chirp'' signal.  The three main types of compact sources of GW interesting for detection purposes are neutron star-neutron star (NS-NS), neutron star-black hole (NS-BH) and black hole-black hole (BH-BH) binaries.  To see NS-NS binaries it is estimated that we need to look to 100-400 Mpc to see three events per year~\cite{Grish,Phin,Narayan,Stairs}.  While it is assumed that there is a greater population of NS-NS binaries, it is the BH-BH binaries which are the strongest candidates for detection from a purely bandwidth point of view.  It is believed that with an initial network of detectors, it will be possible to detect $2\sim3$ events per year at a distance of 200 Mpc~\cite{Grish}.  Due to the masses involved, the inspiral phase for these systems shuts off earlier than in BH-NS or NS-NS systems.  This is advantageous as these binaries should ``chirp'' in or about the most sensitive frequency of the detectors.

As we have some prior knowledge on what the waveform should look like, the optimal strategy is ``Wiener'' or ``matched'' filtering~\cite{Helst}.  This works as follows : we firstly create a set of theoretical models or ``templates'' based on a certain parameter set.  These templates are then cross-correlated with the detector output.  The signal-to-noise ratio (SNR) is compared to a predetermined threshold to decide whether or not a signal is present.  The cross-correlation is weighted with the inverse of the noise power-spectral-density of the detector.  This has the effect of emphasising the most sensitive frequencies of the detector.  This is another reason why the BH-BH binaries will indeed be the best candidates.

In order that the search for GW is not computationally exhaustive we need to have some predetermined criteria for the generation and placement of templates.  In reality a template is dependent on a number of parameters such as mass, spin, eccentricity etc. of the binary.  As the composition of the signal is unknown, the templates have to be laid out in a lattice such that the correlation between a possible signal and a nearby template achieves a predetermined threshold.  In this study, the templates used represent non-spinning circular binaries.  We therefore have a reduced parameter space as we can theoretically model our system by using only the individual masses of the binary components.  It has been shown that for detection purposes it is sufficient to work with {\em restricted post-Newtonian} waveforms~\cite{Cutetal2}.  This is when the amplitude of the waveform is kept at the dominant Newtonian level, but the phase is extended to various post-Newtonian (PN) orders~\cite{Poisson1,Cutetal1,TagNak,Sasaki,TagSas,BDIWW,BDI,WillWise,BIWW,Blan1,TTS}.  While it may still be sufficient to work with restricted waveforms, there may be evidence to suggest that the standard PN waveforms may not be good enough.  

The PN approximation is a perturbative method which expands the equations of motion, binding energy and GW flux as a power series in $v/c$, where $v$ is the orbital velocity and $c$ is the speed of light.  In the early stages of an inspiral, the radiation reaction time-scale, $\tau_{rr}$ is much greater than the orbital time-scale, $\tau_{orb}$.  It is in this adiabatic regime that the PN approximation works best.  The main problem for PN waveforms comes when we enter the strong field regime.  To get an idea of how well the PN waveforms perform, we can look at the ratio $\tau_{rr}/\tau_{orb}$ for various systems in and around the most sensitive frequency of the detector.  If we take the LIGO noise curve, we are mostly interested in the region between $100-300$ Hz.  For a 1.4-1.4 $M_{\odot}$ binary the ratio between the time-scales lies between $4.5\times10^{3}\leq\tau_{rr}/\tau_{orb}\leq 680$.  However, for a 10-10 $M_{\odot}$ binary, the ratio lies between $140\leq\tau_{rr}/\tau_{orb}\leq 40$, where the final value is taken to be the frequency at the last stable orbit ($f_{lso}\approx 190$ Hz).  We know that the adiabatic approximation breaks down when the two time-scales are comparable.  We can see that even for the lowest mass BH-BH binaries expected to be detected, we cannot have much confidence in the performance of the PN templates.  Recently, Damour, Iyer and Sathyaprakash introduced P-approximant waveforms~\cite{DIS1}.  These templates are constructed by taking the PN expansion, defining new energy and flux functions and resumming the series expansion of the new functions using the method of Pad\'e approximation.  These new waveforms were shown to have better behaved convergence properties than the PN templates.  This eliminated the problem with PN waveforms that a higher approximation didn't necessarily mean a better template.

For the placement of templates it has been shown that by incorporating techniques of differential geometry we can greatly simplify the problem~\cite{Owen}.  As the matched filtering is to be carried out in the Fourier domain, it can be shown that the restricted PN Fourier transform waveforms (RPNFT) can be thought of as vectors in a Hilbert space.  Sathyaprakash has shown that when considering two-mass parameter waveforms, a convenient coordinate system to use is not the individual masses but the Newtonian and 1-PN ``chirptimes''~\cite{Sath2}.  The chirptimes are convenient combinations of the total mass, $m = m_{1} + m_{2}$, and the reduced mass ratio, $\eta = m_{1}m_{2}/m^{2}$, of the system.  The concept of a metric tensor on the search manifold was defined by means of two different but ultimately equivalent methods~\cite{Owen,BSD}.  The number of templates needed for a one-step search has already been calculated at Newtonian~\cite{SathDhur}, 1-PN~\cite{Owen} and 2-PN~\cite{OwSath} orders.  In this paper we will extend the previous works in three ways : (i) use the more recent models of the noise power spectral density of the detector~\cite{DIS2} (ii) expand the template number calculation to 2.5-PN for RPNFT waveforms and (iii) calculate the template number for P-approximant waveforms to 2.5-PN order.

The rest of this paper is organized as follows.  In Sec.~\ref{sec:geometry} we recap and partially extend the geometric method of GW detection.  In Sec.~\ref{sec:waveform} we define the T and P-approximant waveforms, while Sec.~\ref{sec:numerical} contains a description of the numerical method used.  Sec.~\ref{sec:results} contains the results for the template number and computational cost using both types of waveforms for all ground-based detectors.

Throughout this paper we use geometric units ($G = c = 1$).

\section{The Geometric Method}\label{sec:geometry}
\noindent The restricted PN time domain waveform can be written as
\begin{equation}
h\left(t, \lambda^{\mu}\right) = a\left(t\right)\,\cos\,\Phi\left(t, \lambda^{\mu}\right).
\label{eq:tdwave2}
\end{equation}
where $\left\{\lambda^{\mu}\right\}$ is an open set on $\Bbb{R}$$^{n}$.  As the search for GW will be carried out in the Fourier domain, it is necessary to have such a representation of the waveform.  In geometric terms the Fourier transform is a bijection from the real space $\Bbb{R}$$^{n}$ to the complex space $\Bbb{C}$$^{n}$.  As the map $h(t, \lambda^{\mu}) \mapsto \tilde{h}(f, \lambda^{\mu})$ is diffeomorphic, this defines a topology on both manifolds~\cite{Prug}.  The phase of the GW in the Fourier domain can be conveniently parameterized by the parameter set $\lambda^{\mu} = \left\{t_{0}, \Phi_{0}, \tau_{k}\right\}$, where $t_{0}$ is the time-of-arrival of the wave, $\Phi_{0}$ is the phase of the wave at time-of-arrival.  The ``chirptimes'',  $\left\{\tau_{k}\right\}$, are functions of the individual masses of the system defined to 1.5-PN approximation by
\begin{equation}
\tau_{0}=\frac{5}{256\eta m^{5/3}\left(\pi f_{0}\right)^{8/3}},  
\label{eq:tau0}
\end{equation}
\begin{equation}
\tau_{2}=\frac{5}{192\eta m\left(\pi f_{0}\right)^{2}}\left(\frac{743}{336}+\frac{11}{4}\eta\right),
\label{eq:tau2}
\end{equation}
\begin{figure}[h]
\begin{center}
\centerline{\epsfxsize=12cm \epsfysize=7cm \epsfbox{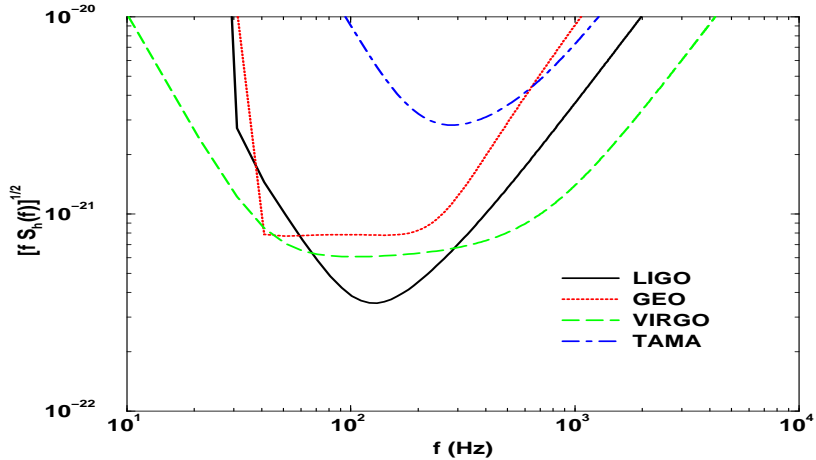}}
\vspace{0.01truecm}
\caption{The one-sided effective noise spectrum for the four ground-based interferometers.}
\label{fig:psd}
\end{center}
\end{figure}
\begin{equation}
\tau_{3}=\frac{\pi}{8\eta m^{2/3}\left(\pi f_{0}\right)^{5/3}} .
\label{eq:tau3}
\end{equation}
The parameter $f_{0}$ corresponds to the lower frequency cutoff of the detector and each subscript, $k$, corresponds to a $(k/2)$-PN approximation.  As we could, in principle, have used the parameter set $\left\{t_{0}, \Phi_{0}, m_{1}, m_{2}\right\}$ to represent the GW phase, we can see than only two of the chirptimes can be independent.  This will prove convenient as it will allow us to choose a combination of any two chirptimes to serve as the coordinate basis for our waveforms.  As the chirptimes are dynamical parameters dependent on the astrophysical system, we will refer to them as being ``intrinsic''.  We will thus refer to the kinematical parameters $t_{0}$ and $\Phi_{0}$ as ``extrinsic'' parameters. 

By treating the Fourier domain waveforms as vectors in a Hilbert space, we can define a scalar product
\begin{equation}
\left<a\left|b\right.\right> =2\int_{0}^{\infty}\frac{df}{S_{h}(f)}\,\,\tilde{a}(f)\tilde{b}^{*}(f) + c.c.,
\label{eq:scalarprod}
\end{equation}
where $c.c.$ denotes complex conjugate, $S_{h}(f)$ is the one-sided noise power spectral density (PSD) and $\tilde{a}(f)$ is the Fourier transform of $a(t)$ defined by
\begin{equation}
\tilde{a}(f) = \int_{-\infty}^{\infty}\,dt\,a(t) e^{2\pi i f t}.
\end{equation}
We can provide an analytical estimate for the shape of the PSD.  In Fig~\ref{fig:psd} we have plotted the effective noise $\sqrt{f\,S_{h}(f)}$ for the four initial ground-based detectors~\cite{DIS2}.  For our purposes, we can think of the factor $1 / S_{h}(f)$ as simply a weight in the scalar product. We can also define a vector norm
\begin{equation}
|\tilde{x}| = \left<x\,|\,x\right>^{1/2}.
\label{eq:norm}
\end{equation}
It is possible to define a metric tensor on the manifold by using two separate, but ultimately equivalent methods.  For both methods we assume that each waveform is normalized to unity. We also consider two points on the manifold separated by the infinitesimal coordinate distance $\Delta\lambda^{\mu}$~\cite{BSD}.  In the first method we use the fact that the proper distance between the two points is given by
\begin{equation}
ds^{2}=|\tilde{h}\left( f,\lambda^{\mu}+\Delta\lambda^{\mu} \right)-\tilde{h}\left( f,\lambda^{\mu}\right) |^{2},
\label{eq:propdist1}
\end{equation}
This can be rewritten as
\begin{equation}
ds^{2}=\left|\frac{\partial \tilde{h}}{\partial\lambda^{\mu}}d\lambda^{\mu}\right|^{2}=\left<\frac{\partial h}{\partial\lambda^{\mu}}\left|\frac{\partial h}{\partial\lambda^{\nu}}\right.\right>d\lambda^{\mu}\,d\lambda^{\nu},
\label{eq:propdist2}
\end{equation}
where the second expression comes from Eqn~(\ref{eq:norm}) after we neglect all terms of order $\Delta\lambda^{2}$ and higher.  The quadratic form on the right hand side suggests we define the metric as
\begin{equation}
\xi_{\mu\nu}=\left<\frac{\partial h}{\partial\lambda^{\mu}}\left|\frac{\partial h}{\partial\lambda^{\nu}}\right.\right>=
2\,\int^{\infty}_{0}\frac{df}{S_{h}(f)}\frac{\partial\tilde{h}^{*}}{\partial\lambda^{\mu}}\frac{\partial\tilde{h}}{\partial\lambda^{\nu}} + c.c.
\label{eq:fullmetric}
\end{equation}
To fully justify this, we can use the fact that in any normed Euclidean vector space, the right hand side of Eqn~(\ref{eq:propdist1}) defines a metric~\cite{Prug,Schutz3,Muk}.  

In the second method we start by calculating the scalar product between two templates $\tilde{h}(\lambda^{\beta})$ and $\tilde{h}(\lambda^{\beta}+\Delta\lambda^{\beta})$ separated by a distance $\Delta\lambda^{\beta}$ in the parameter space.  The scalar product is nothing more than the cosine of the angle between the waveforms.  It has been shown that by expanding the scalar product to quadratic order we obtain~\cite{Owen}
\begin{eqnarray}
\left<\tilde{h}(\lambda^{\beta})\left|\tilde{h}(\lambda^{\beta}+\Delta\lambda^{\beta})\right>\right. & = & 1-\frac{1}{2}\left<\frac{\partial h}{\partial\lambda^{\mu}}\left|\frac{\partial h}{\partial\lambda^{\nu}}\right.\right>\,\Delta\lambda^{\mu}\Delta\lambda^{\nu}, \\
& = & 1 - \frac{1}{2}\xi_{\mu\nu}\,\Delta\lambda^{\mu}\Delta\lambda^{\nu}.
\label{eq:metric2}
\end{eqnarray}
The quantity on the left hand side above is referred to either as the match or overlap between templates.  For this study we will use the term overlap.  We thus rewrite the preceding equation as 
\begin{equation}
O = 1 - \frac{1}{2}\,ds^{2},
\end{equation}
where $O$ denotes the overlap.  As all the parameters of the waveform are not independent, the search for GW will only need to be carried out on the 2-D sub-manifold of independent intrinsic parameters.  In order to project from our initial n-D space to the 2-D sub-space we recursively use a projection onto orthogonal sub-spaces.  This is achieved by using
\begin{equation}
\gamma_{\mu\nu} = \xi_{\mu\nu} - \frac{\xi_{\mu_{0}}\,\xi_{_{0}\nu}}{\xi_{_{00}}}.
\label{eq:projection}
\end{equation}
This projection operation is equivalent to a mapping from the n-D space onto an orthogonal hypersurface.  The projection itself has two separate meanings.  From a geometric point of view, it can be seen as a minimization of distance between two templates.  From a data analysis point of view, it can be viewed upon as a maximization over extrinsic parameters.  By using this projection, we ensure that the overlap between templates on the 2-D sub-space is maximized over all extrinsic parameters.  In Appendix A we outline some of the properties of this projection from both a geometric and detection point of view.

Once we are on the sub-manifold of intrinsic parameters, the number of templates is given by simply dividing the proper volume of the manifold by the volume of the template, $dl^{2}$,  i.e.
\begin{equation}
{\mathcal N} = \frac{\int\int \sqrt{g}\,\,d{\bf\tau}}{dl^{2}} 
\label{eq:templnumbb}
\end{equation}
where $g = det | g_{ij} |$ is the determinant of the metric tensor on the 2-D manifold.  For a rectangular lattice of templates in $N$ dimensions, the volume of each template is given by
\begin{equation}
dl^{N} = \left( 2\sqrt{\frac{1 - {\mathcal MM}}{2}} \right)^{N}.
\end{equation} 
where ${\mathcal MM}$ is the minimal match.  This is a quantity chosen by the experimenter and defines the minimum acceptable loss of event rate.

\section{The Gravitational Waveform}\label{sec:waveform}
The usual restricted PN waveforms are essentially a power series in the velocity $v$, plus logarithmic terms which arise at order $v^{6}$ and beyond.  As we are only considering waveforms up to order $v^{5}$ in this study we can neglect the logarithmic terms and deal with a pure power series.  We can create the Pad\'e approximation by simply equating the power series on the left hand side to a rational function of the right hand side, i.e.
\begin{equation}
\newcommand\Dfrac[2]{\frac{\displaystyle #1}{\displaystyle #2}}
\sum_{k=0}^{n} a_{k}v^{k} = P^{N}_{M} = \Dfrac{\sum_{k=0}^{N} A_{k}\,v^{k}}{\\ 1+\sum_{k=1}^{M} B_{k}\,v^{k}},
\label{eq:ratfunc2}
\end{equation}
such that $n+1 = N + M + 1$.  For a more complete discussion of Pad\'e approximation, we refer the reader to reference~\cite{DIS1}.
 
While it is possible to write the Fourier domain waveform in terms of the chirp masses, in order to use Pad\'e approximation we need to use a more basic format for the wave.  Before progressing, it is necessary to clarify some nomenclature.  The phase of the waveform is normally constructed using a PN-expansion in the velocity, $v$.  As we are dealing with what is in effect a Taylor series, we will call any waveform using PN-approximation a {\em T-approximant} waveform, while any waveform which uses a Pad\'e approximation, we call  a {\em P-approximant} waveform. 

\subsection{T-approximant Waveforms.}\label{subsec:4e1}
\noindent Using the Stationary Phase approximation, we write the Fourier transform of the GW as~\cite{SathDhur,Thorne,DWS,DIS3}
\begin{equation}
\tilde{h}(f) = {\mathcal A} f^{-7/6} \exp\left\{ i \left[ \psi - \frac{\pi}{4}\right] \right\},
\end{equation}
where the phase, $\psi$, is given by the expression
\begin{equation}
\psi = 2 \pi f t_{0} - \phi_{0} + 2 \int_{v_{f}}^{v_{lso}}\,dv \left(v_{f}^{3} - v^{3}\right) \,\frac{E\,'(v)}{F(v)},
\label{eq:newphase}
\end{equation}
where $v_{f} = (\pi m f)^{1/3}$ is the instanteous velocity at frequency $f$, $v_{lso}$ is the velocity at the last stable circular orbit (LSO) and the normalization constant ${\mathcal A}$ is given by demanding $\left<\tilde{h}\,|\,\tilde{h}\right> = 1$.  In the comparable mass case, the orbital energy and GW flux to ${\mathcal O}(v^{6})$ order are given by~\cite{BDIWW,Blan1}
\begin{equation}
E(v) = -\frac{\eta v^{2}}{2}\left[1-\left(\frac{9+\eta}{12} \right)v^{2} - \left(\frac{81-57\eta+\eta^{2}}{24} \right)v^{4} + {\mathcal O}(v^{6})\right].
\label{eq:TaylorE}
\end{equation}
\vspace{1mm}
\begin{eqnarray}
F(v) & = &-\frac{32\eta^{2} v^{10}}{5}\left[1-\left(\frac{1247}{336} + \frac{35}{12}\eta \right)v^{2} + 4\pi v^{3}\right. \nonumber \\ \nonumber\\
&  &\left. + \left(-\frac{44711}{9072} + \frac{9271}{504}\eta + \frac{65}{18}\eta^{2}\right)v^{4} - \left(\frac{8191}{672} + \frac{535}{24}\eta \right)\pi v^{5} + {\mathcal O}(v^{6})\right].
\label{eq:TaylorF}
\end{eqnarray}
The position of the LSO is unknown in the case of two comparable mass objects.  However, it has been hypothesized that a more accurate description of $v_{lso}$ may be given by~\cite{DIS1} 
\begin{equation}
v_{lso}^{2} = \frac{1}{6}\frac{1+\frac{1}{3} \eta}{1-\frac{35}{36}\eta}\left(2-\frac{1+\frac{1}{3} \eta}{\sqrt{1-\frac{9}{6}\eta+\frac{1}{36}\eta^{2}}}\right).
\end{equation}
which approaches the Schwarzchild test-mass limit in the case of $\eta\rightarrow0$.  This in turn should provide more accurate knowledge of the position of the LSO.  From this we can define the frequency at the LSO as $f_{lso} = v_{lso}^{3} / \pi m$.

\subsection{P-approximant Waveforms.}\label{subsec:4e2}
The groundwork for constructing the Pad\'e approximation of the energy and flux has already been laid down~\cite{DIS1}.  In this section we will briefly outline the main points.  As we are dealing with only sub-diagonal Pad\'e approximants we can denote the rational function $P^{N}_{M}$ by $P^{m}_{m+\epsilon}$, where as we increase the level of approximation the parameter $\epsilon = 0$ or 1.  The advantage of the sub-diagonal approximants is purely computational in that we can cast the flux in a continued fraction form and we only have one new coefficient to calculate as we go to each next level of approximation.  

\subsubsection{The P-approximant Energy Function}
Starting with Eqn~(\ref{eq:TaylorE}) we can construct the Pad\'e approximant energy function
\begin{equation}
E_{P_{n}}(x) = \left[ 1+2\,\eta\left(\sqrt{1+e_{P_{n}}(x)}-1\right)\right]^{1/2}-1,
\label{eq:newEnergy}
\end{equation}
where $x=v^{2}$ and $e_{P_{n}}(x)$ is an energy flux related to the binding energy of the system given by
\begin{equation}
e_{P_{n}}(x) = -x\,P^{m}_{m+\epsilon}\left[\sum_{k=0}^{n}a_{k}\,x^{k}\right],
\end{equation}
where the coefficients $a_{k}$ are given by
\begin{equation}
a_{0} = 1, \,\,\,\,\,a_{1} = -1-\frac{\eta}{3}, \,\,\,\,\,a_{2} = -3 + \frac{35\eta}{12}.
\end{equation}
Constructing the Pad\'e approximation gives us the expression
\begin{equation}
e_{P_{n}}(x) = -x\,\frac{1+\frac{1}{3}\eta - \left(4 - \frac{9}{4}\eta + \frac{1}{9}\eta^{2}\right)x}{1 + \frac{1}{3}\eta - \left(3 - \frac{35}{12}\eta \right)x}.
\label{eq:newenergy}
\end{equation}
Once again, taking the derivative of Eqn~(\ref{eq:newEnergy}) with respect to the velocity yields
\begin{equation}
E^{\,\,\prime}_{P_{n}}(v) = \frac{v\eta}{\left(1 + E_{P_{n}}(x)\right)\sqrt{1 + e_{P_{n}}(x)}}\frac{de_{P_{n}}(x)}{dx},
\end{equation}
where
\begin{equation}
\frac{de_{P_{n}}(x)}{dx} = -c_{0}\frac{1 + c_{2}x\left(2 + \left(c_{1} + c_{2} \right) x\right)}{\left(1 + \left(c_{1} + c_{2} \right) x\right)^{2}},
\end{equation}
and the Pad\'e coefficients, $c_{k}$, are given by
\begin{equation}
c_{0} = 1, \,\,\,\,\,\,c_{1} = 1 + \frac{\eta}{3}, \,\,\,\,\,\,c_{2} = -\frac{4 - \frac{9}{4}\eta + \frac{1}{9}\eta^{2}}{1 + \frac{1}{3}\eta}.
\end{equation}

\subsubsection{The P-approximant Flux}
As we are only dealing with waveforms up to 2.5-PN order and can neglect all logarithmic terms, we can simplify the steps taken for converting the flux into a form compatible with Pad\'e theory.  Starting with Eqn~(\ref{eq:TaylorF}) we can write the PN flux as
\begin{equation}
F_{T_{n}}(v) = F_{N}(v)\sum_{k = 0}^{5} c_{_{k}}\,v^{k}, 
\end{equation}
where $F_{N}(v)$ is the Newtonian flux given by
\begin{equation}
F_{N}(v) = \frac{32}{5}\eta^{2}v^{10}.
\end{equation}
With this representation of the flux, the idea is to create a three step mapping to a Pad\'e approximant flux, $F_{P}(v)$.  In order to use Pad\'e theory, our power series needs to contain a linear term in the velocity $v$.  The reason for this is that the linear coefficient of the Taylor series appears in the denominator of the Pad\'e coefficients.  A Taylor series like the PN-flux, which contains a null linear term, results in infinite Pad\'e coefficients.  The first step therefore is to introduce such a term and create a ``factored'' flux function by the operation
\begin{eqnarray}
f_{T_{n}} & = & \left(1-\frac{v}{v_{pole}}\right)F_{T_{n}},\\
          & = & F_{N}(v)\,\sum_{k = 0}^{5} f_{_{k}}\,v^{k}.  
\end{eqnarray}
where $v_{pole}$ is the velocity at the photon ring, $f_{_{0}} = c_{_{0}}$ and $f_{_{k}} = c_{_{k}} - c_{_{k-1}}/v_{pole}$ from $k = 1,..,n$.  This choice of using $v_{pole}$ as a normalizing velocity comes naturally from the fact that we expect a pole at $r = 3\,M$ in the flux for particles in a Schwarzchild metric.  The next step is to create the corresponding P-approximant ``factored'' flux by 
\begin{equation}
f_{P_{n}}(v) = F_{N}(v)\,P^{m}_{m+\epsilon}\left[\sum_{k=0}^{n}\,f_{_{k}}v^{k}\right].
\end{equation}
The final P-approximant flux is then given by
\begin{equation}
F_{P_{n}}(v) = \left(1-\frac{v}{v_{pole}}\right)^{-1}\,f_{P_{n}}(v).
\end{equation}
It has been shown that this representation of the flux has faster convergence properties than the usual PN-flux.

\subsection{Comparison between T and P-approximant Waveforms.}
The question must be asked as to why we should favor one template model over another.  While both templates are based on the adiabatic approximation, it has been shown that the P-approximant templates are superior in a number of ways~\cite{DIS1,DIS2}.  The main problem with the T-approximant waveforms is that as we go to higher approximations, we do not necessarily obtain a better template.  In the test-mass approximation, we know that the $v^{4}$ and $v^{5}$ PN approximants diverge in opposite directions from the exact numerical flux.  This is reflected when we go to the comparable mass case in the values of the overlaps between the 2 and 2.5-PN waveforms and the fiducial ``exact'' waveform.  For a 10-10 $M_{\odot}$ system, the 2-PN waveform achieves an overlap of 0.894, while the 2.5-PN template has an overlap of 0.545.  Compare this with the results of the P-approximant waveforms : 0.868 for the 2-PN template and 0.979 for the 2.5-PN template, and we can already see the benefit of the P-approximant waveforms.  Another test can be carried out by looking at the Cauchy convergence between the approximations.  If we compare the 2 and 2.5-PN approximations for both types of waveforms, we get a value of 0.496 between the T-approximant waveforms and a value of 0.918 for the P-approximant waveforms. 

The next problem encountered with T-approximant waveforms is that it {\em is} possible to achieve overlaps of $\geq$ 0.9.  The price that has to be paid, however, is a large bias in the parameter values.  For example it has been shown that for a [1.4,10] $M_{\odot}$ system, it is possible to achieve an overlap of 0.94 with a 2.5-PN template~\cite{DIS2}.  However, this is achieved using a template with masses of 0.82 and 20.4 $M_{\odot}$.  In comparison, the P-approximant template achieves a higher overlap (0.994) with lower biases in the parameter values ([1.37, 10.02] $M_{\odot}$).  From this we must conclude that the PN waveforms are not faithful templates.

Taking into account how badly the PN templates perform in the test mass case, we should be worried about their performance in the case of two comparable masses.  On the other hand, it is due to their faster convergence properties and lower biases in parameter estimation, why we believe that the P-approximant templates will give superior results in the search for gravitational waves.

\section{The Numerical Method}\label{sec:numerical}
\begin{figure}[h]
\begin{center}
\centerline{\epsfxsize=12cm \epsfysize=7cm \epsfbox{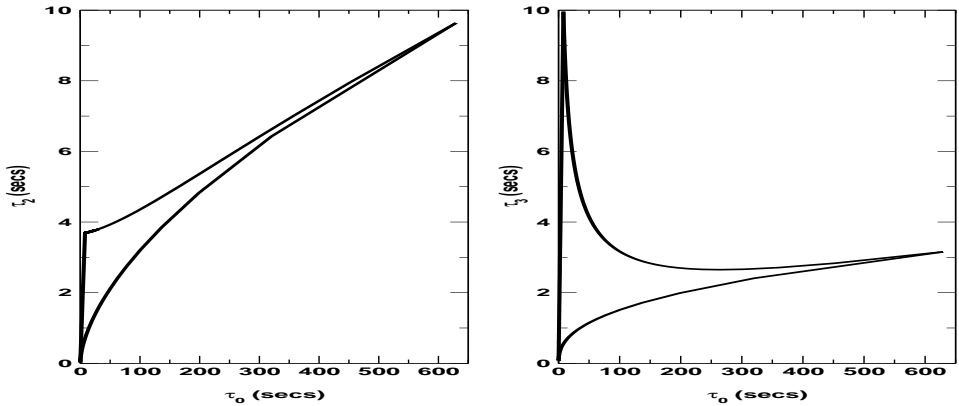}}
\vspace{0.001truecm}
\caption{The $\tau_{0}$-$\tau_{2}$ and $\tau_{0}$-$\tau_{3}$ parameter spaces for a range of total masses $m \in [ 0.4 , 100 ] M_{\odot}$.  The range of astrophysically interesting sources is contained in both cases within the wedge.}
\label{fig:paramspace}
\end{center}
\end{figure}

\noindent Our primary aim in this study is to calculate the number of templates needed for a GW search using P-approximant waveforms.   This is closely followed with the aim of developing a numerical code which can be used for any type of waveform.  As the calculation of the metric tensor depends on the derivatives of the waveforms, it is sometimes more convenient to do this numerically.  The eliminates the need for calculating the derivatives analytically, which for example, in the case of P-approximant waveforms is very messy.  In order to calculate the numerator of Eqn~(\ref{eq:templnumbb}) there are a number of possible methods.  Due to the shapes of the parameter spaces (Fig~\ref{fig:paramspace}), we chose to use a two-fold Monte Carlo method for evaluating the double integral in the same equation.  This method is a combination of two different Monte Carlo methods called ``hit-or-miss'' and ``crude'' methods~\cite{NumRec,HamHan,Raj}. The ``hit-or-miss'' method is used
\noindent for calculating the fractional area of the parameter space, $\delta A$.  Briefly, this involves enclosing the parameter space inside a boundary of known area, and throwing points at random inside the boundary.  The fractional area is then given by $\delta A = A \times (n / N)$, where $A$ is the area of the known boundary, $n$ is the number of points falling inside the region of interest and $N$ is the total number of points.  If the search manifold was Euclidean we would only need to calculate the value $\sqrt{g}$ once.  For a toy model where all waveforms are terminated at a constant frequency irregardless of mass, then it has been shown that the manifold is approximately Euclidean~\cite{TanTag}.  A suitable co-ordinate transformation maps the manifold onto a globally Euclidean space where the metric is constant.  However, as soon as we include the frequency at the LSO, $f_{lso}$, in the termination of the waveforms, the search manifold becomes a curved Riemannian manifold and a globally Euclidean manifold no longer exist~\cite{PB}.  This is where the ``crude'' method comes in.  We can us it to calculate the mean value of $\sqrt{g}$ on the manifold.  By using the chirptimes introduced earlier as coordinate basis, we have a choice of two coordinate systems within which to work.  These are the $\tau_{0}$-$\tau_{2}$ and $\tau_{0}$-$\tau_{3}$ systems.  What we found, from a numerical point of view, was that the fractional area $\delta A$ converged quickest in the $\tau_{0}$-$\tau_{2}$ coordinate system. From a computational point of view, this makes it a more desirable set of coordinates as it decreases the number of necessary calculations.

\noindent When generating numerical waveforms, the Nyquist theorem states that for a bandwidth limited signal with a critical frequency, $f_{Nyq}$, then in reconstructing the time series waveform our sampling rate, $f_{samp}$, must be at least twice the critical frequency.  This is to ensure that any frequency above the critical frequency will not lead to an 'aliasing' in the waveform.  This is when higher frequency components of the wave occur within the required bandwidth.  As previously stated, in this investigation it was decided to evolve the waveforms as far as $f_{lso}$ or to some arbitrary cutoff frequency.  The choice is made as follows.  When using the arbitrary cutoff, a frequency of $f_{cut} = 2\,kHz$ is chosen. This means that all binaries with total masses in the range $m \in [2.1,100]\, M_{\odot}$  are evolved up to and beyond their natural cutoff frequencies (by this it is assumed that the natural cutoff frequency for an inspiral waveform is  at the last stable circular orbit), while only the lowest mass binaries have their waveforms cut off early. It is not expected that this will have any great effect on the number of templates as $99\%$ of the waves energy is extracted by the time the GW frequency has reached $750\,Hz$~\cite{BSD}. This can easily be seen from looking at the power spectrum of $\tilde{h}(f)$ which is proportional to $f^{-7/3}$.   

To ensure there is no aliasing in the construction of our waveforms we have chosen the sampling frequency to be $f_{samp} = 6\,kHz$ if $f_{lso} > 1.5\,kHz$,  or, $f_{samp} = 4\,f_{lso}$ if $f_{lso} \leq 1.5\,kHz$.  As our waveforms have a maximum bandwidth of $2\,kHz$, our sampling frequency is always at least three times $f_{Nyq}$.  To cut down on computation time we are starting out however with the Fourier transforms rather than the time series waveforms.  

In order to investigate the reliability of the numerical code, we tested it against known results calculated semi-analytically.  These results were calculated previously for RPNFT waveforms up to 2-PN order using older noise curves~\cite{OwSath}.  The basic assumptions of that study were as follows :  a rectangular template lattice was used with a minimal match of 0.97 and a SNR of 1.  It was assumed that the minimal individual mass was $0.2\,M_{\odot}$. Using these same assumptions, we calculated the template number for identical waveforms using the numerical method outlined above.  The results from the numerical code coincided with previous results to within $5\%$.    Thus, we can be confident about our results for template number for both the T and P-approximant waveforms using the new noise curves

\section{Results and Discussion}\label{sec:results}
For the initial ground-based interferometers the most likely sources to be detected are solar-mass BH-BH binary systems.  This is because these systems have an LSO frequency in or around the most sensitive points of the detector noise curves and are therefore the best candidates for detection.  For completeness, however, we would also like to know how many templates are needed to search for BH-NS and NS-NS binaries as well.  With this in mind, we calculated the number of templates for two separate cases.  In the first case we chose a minimum individual mass of $1\, M_{\odot}$, while in the second we chose $5\, M_{\odot}$.  The maximum total mass in each case is different due to the different lower frequency cutoffs of the detectors.  The values used are $86\,M_{\odot}$ (LIGO, GEO), $100\,M_{\odot}$ (VIRGO) and $50\,M_{\odot}$ (TAMA).  For both cases we used the same assumptions. Namely, we choose a minimal match of ${\mathcal MM} = 0.97$.  This means that we have a loss in detection rate loss $\sim10\%$.  All waveforms are terminated at the LSO frequency or $2\,kHz$, whichever is reached first.  And finally we choose a rectangular lattice of templates.  
  
As well as template number, we need to know the computational cost of the storage of these templates.  Schutz~\cite{Schutz2} has provided a relation between the number of templates and the CPU requirements for a GW search.  In brief terms, the data streams from the detector are broken into segments of $D$ real numbers.  The length of the longest filter is $F$ real numbers, such that $F>>D$.  To filter the data through ${\mathcal N}$ templates of length $F$ requires ${\mathcal N}\,D\left(16 + 3 \log_{2}F \right)$ floating point operations (flops).  If the sampling frequency is $2\,f_{Nyq}$, then the computational power required is
\begin{equation}
P\approx {\mathcal N} f_{Nyq} \left(32 + 6 \log_{2}F \right)
\end{equation}
flops.  If we assume that the length of the longest filter is due to a minimal mass template such that $m_{1}^{min} + m_{2}^{min} = m^{min}$, $\eta = 0.25$, then the length of the longest filter, $F$, is given by
\begin{equation}
F = \frac{5}{32}\,f_{Nyq} \left(\pi f_{low}\right)^{-8/3}\left(2 m^{min}\right)^{-5/3}
\end{equation}
where we once again assume a sampling rate of $2 f_{Nyq}$.  These equations can be amended accordingly depending on the sampling rate used.

\begin{figure}[t]
\begin{center}
\centerline{\epsfxsize=11cm \epsfysize=9cm \epsffile{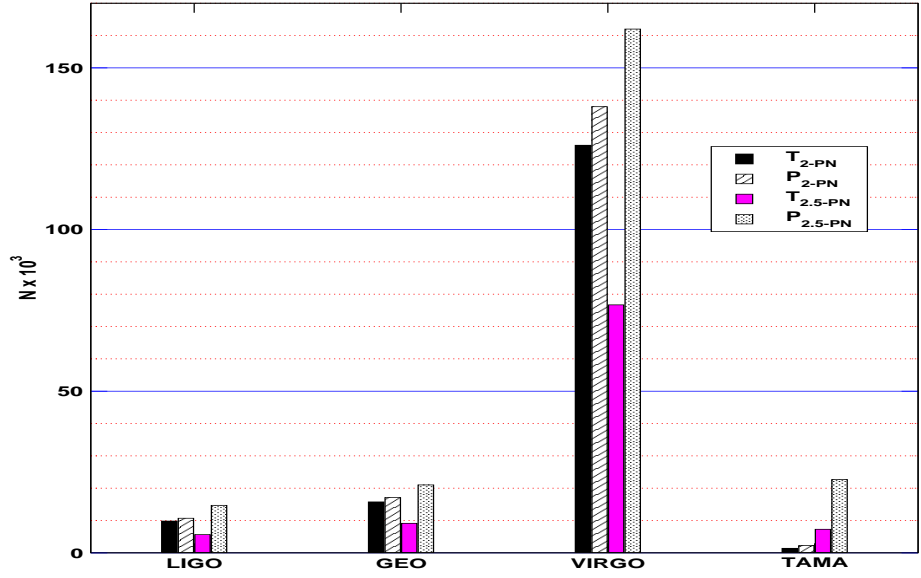}}
\vspace{0.05mm}
\caption{A comparison of the number of templates needed at 2 and 2.5-PN orders for a GW search using T and P-approximant waveforms assuming a minimal match of 0.97 and a minimum total mass of $2\,M_{\odot}$.}
\label{fig:barTP1}
\end{center}
\end{figure}

\begin{figure}[b]
\begin{center}
\centerline{\epsfxsize=11cm \epsfysize=9cm \epsfbox{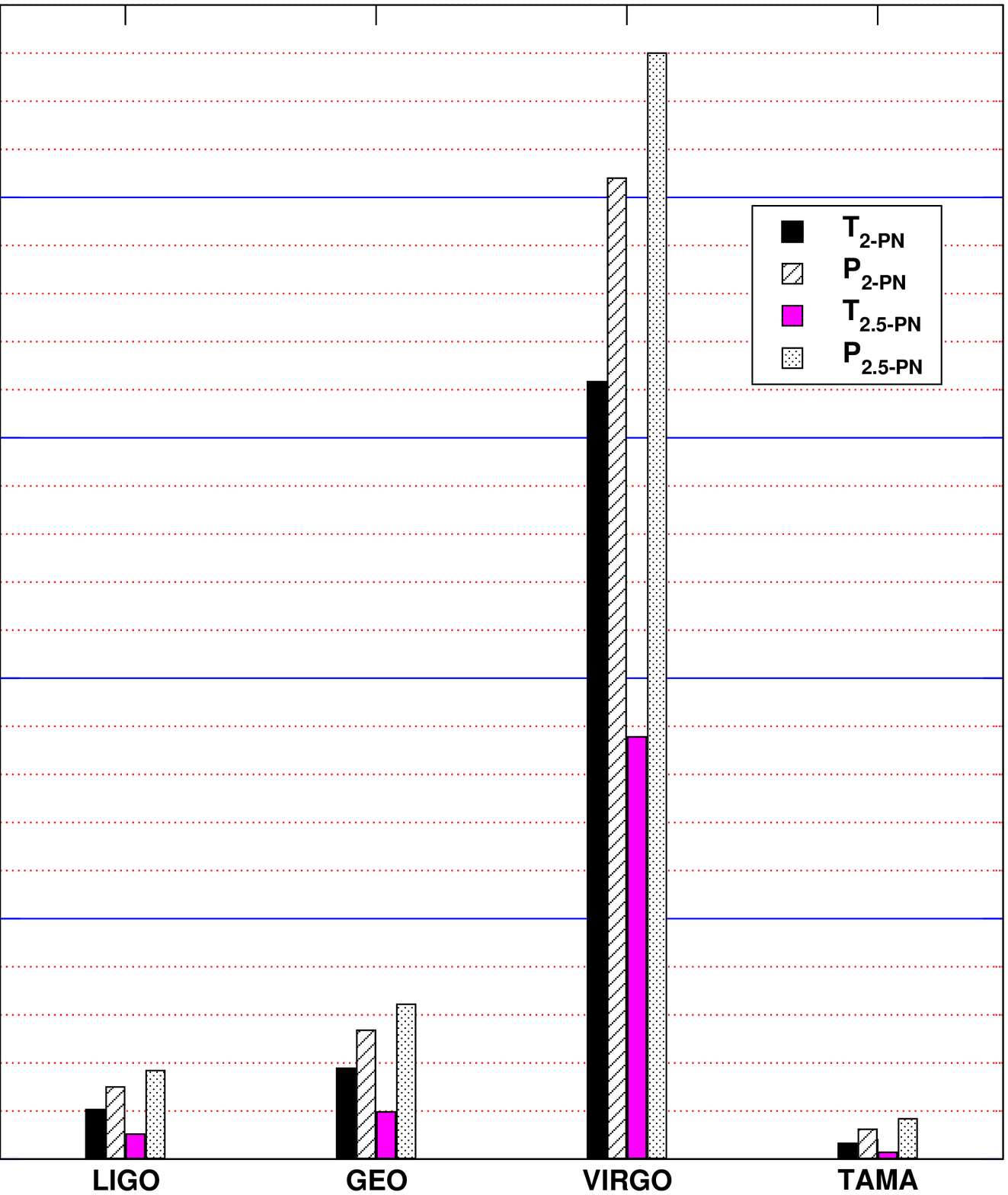}}
\vspace{0.05mm}
\caption{A comparison of the number of templates needed at 2 and 2.5-PN orders for a GW search using T and P-approximant waveforms assuming a minimal match of 0.97 and a minimum total mass of $10\,M_{\odot}$.}
\label{fig:barTP5}
\end{center}
\end{figure}

\begin{figure}[h]
\begin{center}
\centerline{\epsfxsize=11cm \epsfysize=9cm \epsfbox{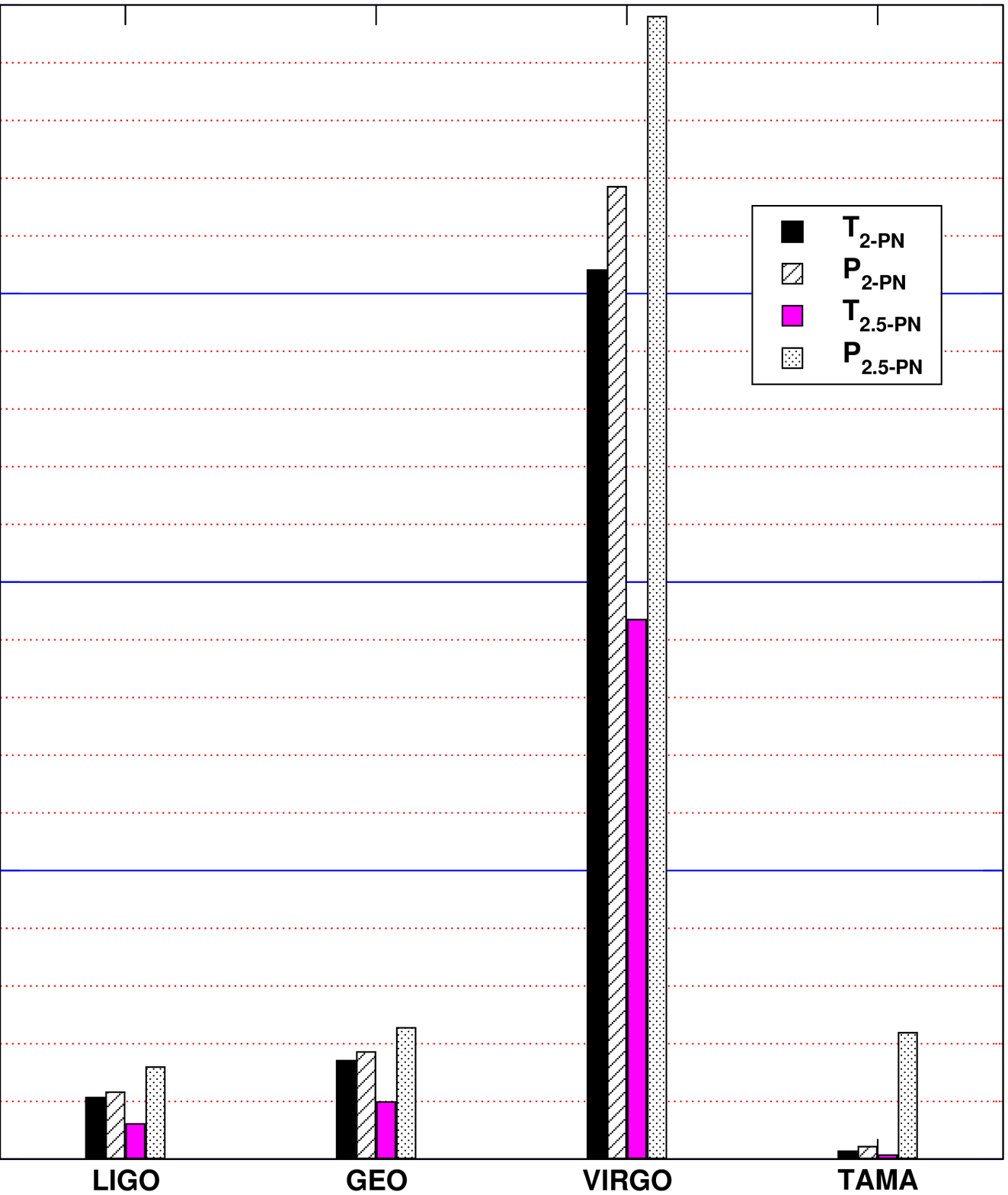}}
\vspace{0.05mm}
\caption{A comparison of the computational cost at 2 and 2.5-PN orders for a GW search using T and P-approximant waveforms assuming a minimal match of 0.97 and a minimum total mass of $2\,M_{\odot}$.}
\label{fig:cpuTP1}
\end{center}
\end{figure}

\begin{figure}[h]
\begin{center}
\centerline{\epsfxsize=11cm \epsfysize=9cm \epsfbox{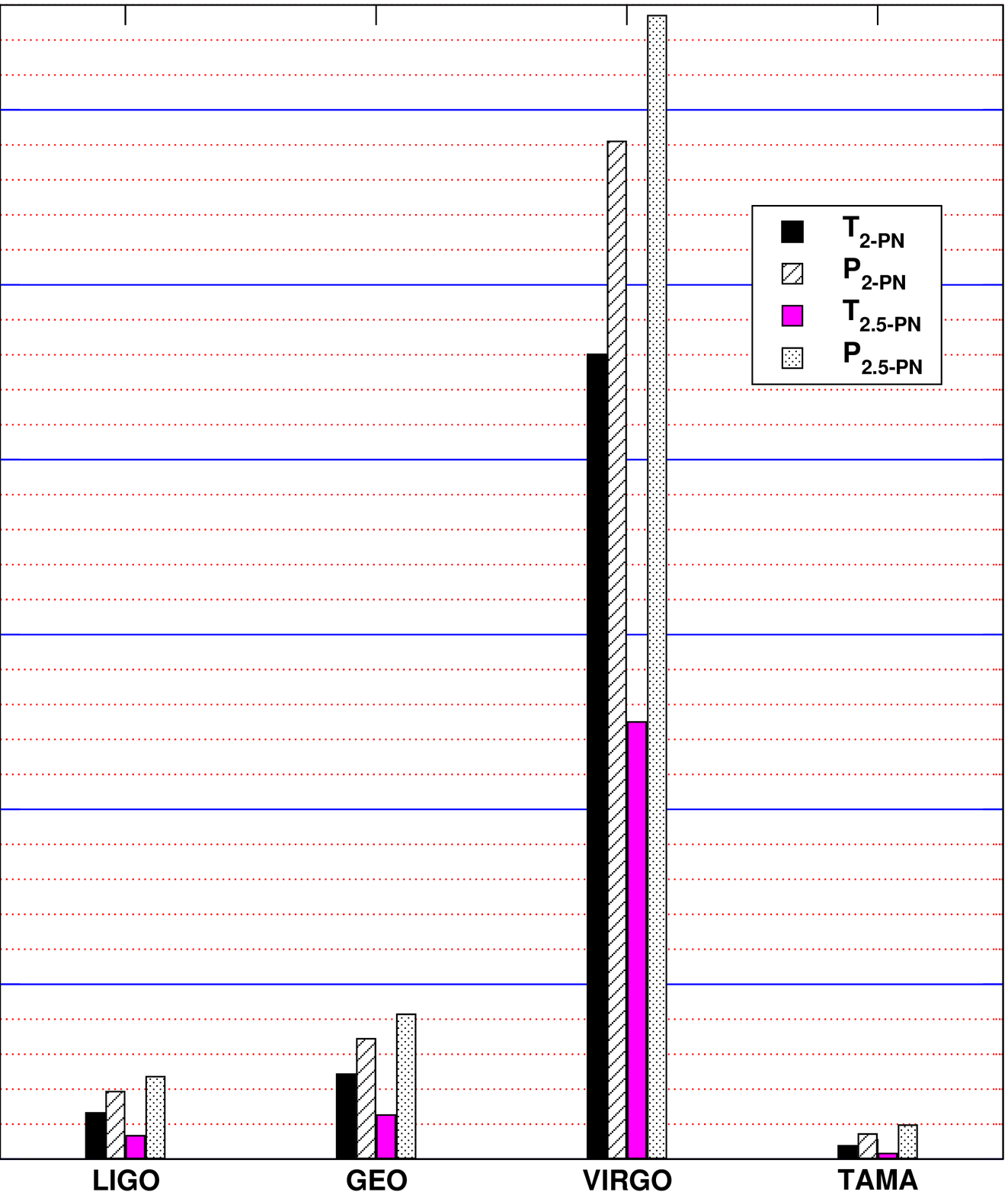}}
\vspace{0.05mm}
\caption{A comparison of the computational cost at 2 and 2.5-PN orders for a GW search using T and P-approximant waveforms assuming a minimal match of 0.97 and a minimum total mass of $10\,M_{\odot}$.}
\label{fig:cpuTP5}
\end{center}
\end{figure}

The number of templates needed for a one-step GW search using 2 and 2.5-PN waveforms are presented in Figs~\ref{fig:barTP1}~-~\ref{fig:barTP5}, while the computational cost is given in Figs~\ref{fig:cpuTP1}~-~\ref{fig:cpuTP5}.  We present the results for these particular approximations as we believe that they currently represent the best templates for a GW search.  We can see by comparing between sets of results, that in most cases we need more templates to search for inspiralling binaries using P-approximant waveforms than we do using T-approximant waveforms.  Initially, this may seem counter-intuitive.  We would normally assume, that if the P-approximant waveforms do indeed bare a closer resemblance to the true waveform of a signal, we would need a lower number of templates.  We thus need to explain the large increase in the number of templates as we go from T to P-approximant waveforms.

\begin{figure}[h]
\begin{center}
\centerline{\epsfxsize=14cm \epsfysize=10cm \epsfbox{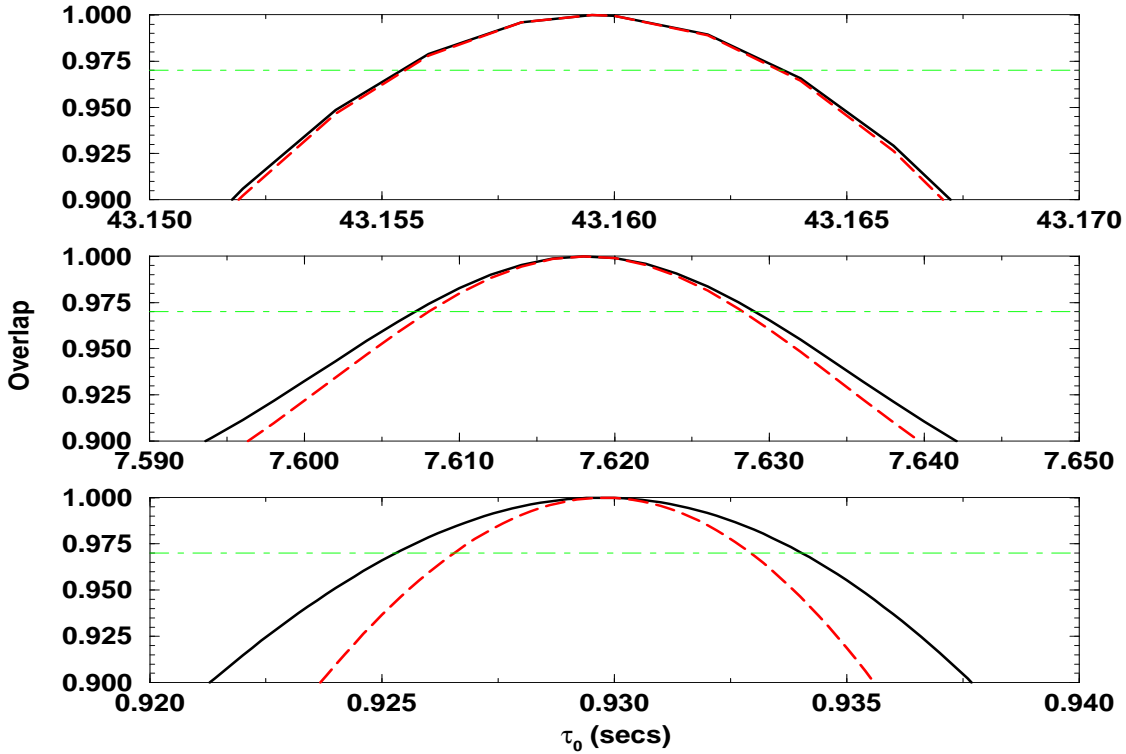}}
\vspace{0.5truecm}
\caption{The span of the T-approximant (solid) and P-approximant templates (dashed) on the $\tau_{0}$ axis.  From top to bottom, the range of masses is $\left[1, 1\right]\,M_{\odot}$, $\left[1, 10\right]\,M_{\odot}$ and $\left[10, 10\right]\,M_{\odot}$.  The dot-dashed line corresponds to an overlap of 0.97}  
\label{fig:span}
\end{center}
\end{figure}

In order to show why we need more templates using P-approximant waveforms, we carried out two experiments.  In the first experiment we use the fact that the extrinsic components of the metric remain constant over the entire manifold.  It is only the intrinsic components that change.  We therefore held all of the waveform parameters constant except for either $\tau_{0}$ of $\tau_{2}$.  We chose a number of different points on the search manifold and varied the intrinsic parameters by $10^{-5}$ seconds.  We then calculated the proper distance between both templates.  We know from Eqn~(\ref{eq:metric2}) that there is a connection between the overlap of two nearby templates and the proper distance between them.  To first order approximation in $\Delta\lambda^{\mu}$, the proper distance is given by
\begin{equation}
s\approx \sqrt{1 - \left<\tilde{h}(\lambda^{\mu})\left|\tilde{h}(\lambda^{\mu}+\Delta\lambda^{\mu})\right>\right.} 
\end{equation}
What we find is that the proper distance between two P-approximant templates is greater than the proper distance between two T-approximant templates at constant points on the manifold.  This is explained by the fact that not only does the proper area of the manifold change as we go to successive approximations, but it also changes as we go from one type waveform to another.  This can be seen from examining the eigenvalues of the metric tensor as we firstly go from one approximation to another, and secondly, by using different types of waveforms.  For the second experiment, we calculated the span of each template at different points on the manifold.  We chose a number of different mass ranges on the manifold.  In Fig~\ref{fig:span} we present the results for the choice of $\left[1, 1\right]\,M_{\odot}$, $\left[1, 10\right]\,M_{\odot}$ and $\left[10, 10\right]\,M_{\odot}$ while changing the Newtonian chirptime $\tau_{0}$.  The solid curve represents the span of the T-approximant template, while the dashed curve represents the span of the P-approximant template.  We can see in the case of $\left[1, 1\right]\,M_{\odot}$, there is not much difference in the span of the templates when using either of the approximations.  However, as we move to higher and higher masses, the P-approximant templates have a smaller span on the search manifold than the T-approximant templates.  Once again, this can be explained by looking at the eigenvalues of the metric.  The orientation and span of each template is dependent on the eigenvalues of the locally diagonal metric.  We can deduce from this that it is in the high mass range of the manifold that is mainly responsible for the change in proper area as we go from T to P-approximant waveforms.  This fact, combined with the orientation of the template, sufficiently explains the need for extra templates when using the P-approximant waveforms.  The numerical results are presented in Appendixes B and C.

We should not be discouraged by the vast increase in templates for P-approximant waveforms at certain approximations.  As we know that the 2.5-PN P-approximant template is vastly superior to its T-approximant counterpart, what is most important is the comparison between 2-PN templates.  In both cases that we studied, there is only a small increase in template number needed as we go from T-approximant to P-approximant waveforms.  This increase is more than acceptable due to the fact that the P-approximant templates are more robust and reliable in general.

\clearpage
\vspace{-5cm}

\clearpage 

\section{conclusion}
In this study we have calculated the number of templates needed for a one-step GW search using T-approximant and P-approximant waveforms.  We first displayed how the detection of GW can be examined by evoking pre-established techniques of differential geometry.  By treating the waveforms as vectors in a Hilbert space we can define an N-dimensional Riemannian manifold upon which we can define a scalar product, vector norm and a metric tensor.  By choosing a particular parameterization for the phase of the GW we showed how the search for GW need only take place on a 2-D submanifold defined by a set of independent parameters intrinsic to the binary system in question.  The operation of projecting onto this sub-manifold has a two-fold interpretation.  From a geometric point of view it is a minimization of distance between two templates, while from a data analysis point of view it is a maximization of parameters.  We have also outlined the development of an easily modifiable  numerical code for calculating template number.  This code uses the latest analytical models of the noise spectral density of the initial ground-based detectors. 

This code uses a combination of ``crude'' and ``hit-or-miss'' Monte Carlo methods to calculate the proper area of the search parameter space.  We found that there was a particular coordinate system ($\tau_{0}$-$\tau_{2}$) which seemed to be particularly advantageous from a computational point of view.  At present the code works for waveforms up to 2.5-PN order.  We outlined the difference between the T and P-approximant comparable mass waveforms, as well as how to construct the P-approximant energy and flux functions.   These waveforms were generated to the LSO for waveforms with $f_{lso} \leq 1.5\, kHz$.  For waveforms with $f_{lso} > 1.5\, kHz$ we introduced a cutoff frequency of $f_{cut} = 2\, kHz$.  This is entirely consistent as most of the waves energy has been extracted by $\sim 1 kHz$.  Although it is BH-BH binaries that present the best possibility for detection by the first generation of detectors, we calculated template number for two seperate cases assuming minimum individual masses of $1\,M_{\odot}$ and $5\,M_{\odot}$

We found that in general we need more templates when using P-approximant waveforms.  In some cases (e.g. 1.5-PN order) the number of templates needed increased by a factor of 3.  However, the 1.5-PN templates are known to be unreliable.  The case we were most interested in was the comparison in performance of the 2-PN templates.  We found that we while we still need a greater number of templates, this increase in small, and when weighed against the known robustness of the P-approximant templates, the increase is acceptable.  We were able to explain the increase in template number by the fact that the proper area of the parameter space is dependent on the level of approximation and type of waveform used.  We used this fact to explain how the span of the P-approximant templates decreases faster than the corresponding T-approximant templates as we increase the total mass.

These results are encouraging for the use of P-approximant templates over the usual PN waveforms in the search for GW.

\section*{Acknowledgements}
I would like to thank Dr B. S. Sathyaprakash for proposing the problem to me and for his invaluable guidance during the research and in the writing of this paper.  I would also like to thank Dr(s) Stanislav Babak, Kostas Glampedakis, Andreas Dimitropoulos, R. Balasubramanian and Prof. L.P. Grishchuk for the extremely useful discussions and suggestions during the period of this work.  I would finally like to thank Phil Fayers for his help and assistance with the computational issues of this problem.

\appendix

\section{Properties of the Metric Projection}

We have defined the metric projection as
\begin{equation}
\gamma_{\mu\nu} = \xi_{\mu\nu} - \frac{\xi_{\mu\,_{0}}\,\xi_{_{0}\,\nu}}{\xi_{_{0\,0}}},
\label{eq:projection1}
\end{equation}
What we will now show is that there is a two-fold interpretation of the projection depending on whether one is looking at the problem from a geometrical or detection point of view.  As one follows on closely from the other, we will first examine the meaning of the projection from a geometrical point of view.
We firstly define the distance between two points (or templates) on a $4-D$ manifold as
\begin{equation}
ds^{2} = \xi_{\mu\nu}\, \Delta\lambda^{\mu}\Delta\lambda^{\nu},
\end{equation}
\begin{equation}
=\xi_{\,_{00}}\,\Delta\lambda^{0}\Delta\lambda^{0} + 2\,\xi_{\,_{0j}}\,\Delta\lambda^{0}\Delta\lambda^{j} + \xi_{\,_{ij}}\,\Delta\lambda^{i}\Delta\lambda^{j},  
\end{equation}
where we use arbitrary coordinates $\lambda^{\mu} = \left\{x, y, z, w\right\}$.  Any function is minimized or maximized if its first derivative is zero.  Therefore, the proper distance, $ds^{2}$, is minimized if
\begin{equation}
\frac{\partial}{\partial\,\Delta\lambda^{\mu}}\left(ds^{2}\right) = 0.
\label{eq:propdistderiv}
\end{equation}
Initially we want to project from an $n-D$ space to an $m-D$ orthogonal sub-space ($m < n$).  We can project onto the orthogonal sub-space as follows.  We first choose to minimize the proper distance with respect to the $x$-coordinate. Therefore, taking the first derivative with respect to $\Delta\lambda^{0}$, where $\Delta\lambda^{0} = \left| x - x_{0}\right|$
\begin{equation}
\Rightarrow \frac{\partial}{\partial\,\Delta\lambda^{0}}\left(ds^{2}\right) = \xi_{\,_{00}}\,\Delta\lambda^{0} + 2\,\xi_{\,_{0j}}\,\Delta\lambda^{j} = 0,
\label{eq:constr1}
\end{equation}
The trivial solution is if $\Delta\lambda^{0} = \Delta\lambda^{j} = 0$.  Therefore, Eqn~(\ref{eq:constr1}) implies
\begin{equation}
\Rightarrow \xi_{\,_{00}} = \xi_{\,_{0j}} = 0.
\end{equation}
The proper distance between points on the $n-1$ sub-manifold is then given by
\begin{equation}
d\sigma^{2} = \gamma_{\mu\nu}\, \Delta\lambda^{\mu}\Delta\lambda^{\nu},
\end{equation}
where $\gamma_{\mu\nu}$ is the metric tensor on the $3-D$ submanifold defined by Eqn~(\ref{eq:projection1}) and $d\sigma^{2} < ds^{2}$.  We can therefore see how the projection of the metric can be interpreted as a minimization of distance between points on the manifold.
\begin{figure}
\begin{center}
\centerline{\epsfxsize=11cm \epsfysize=7cm \epsfbox{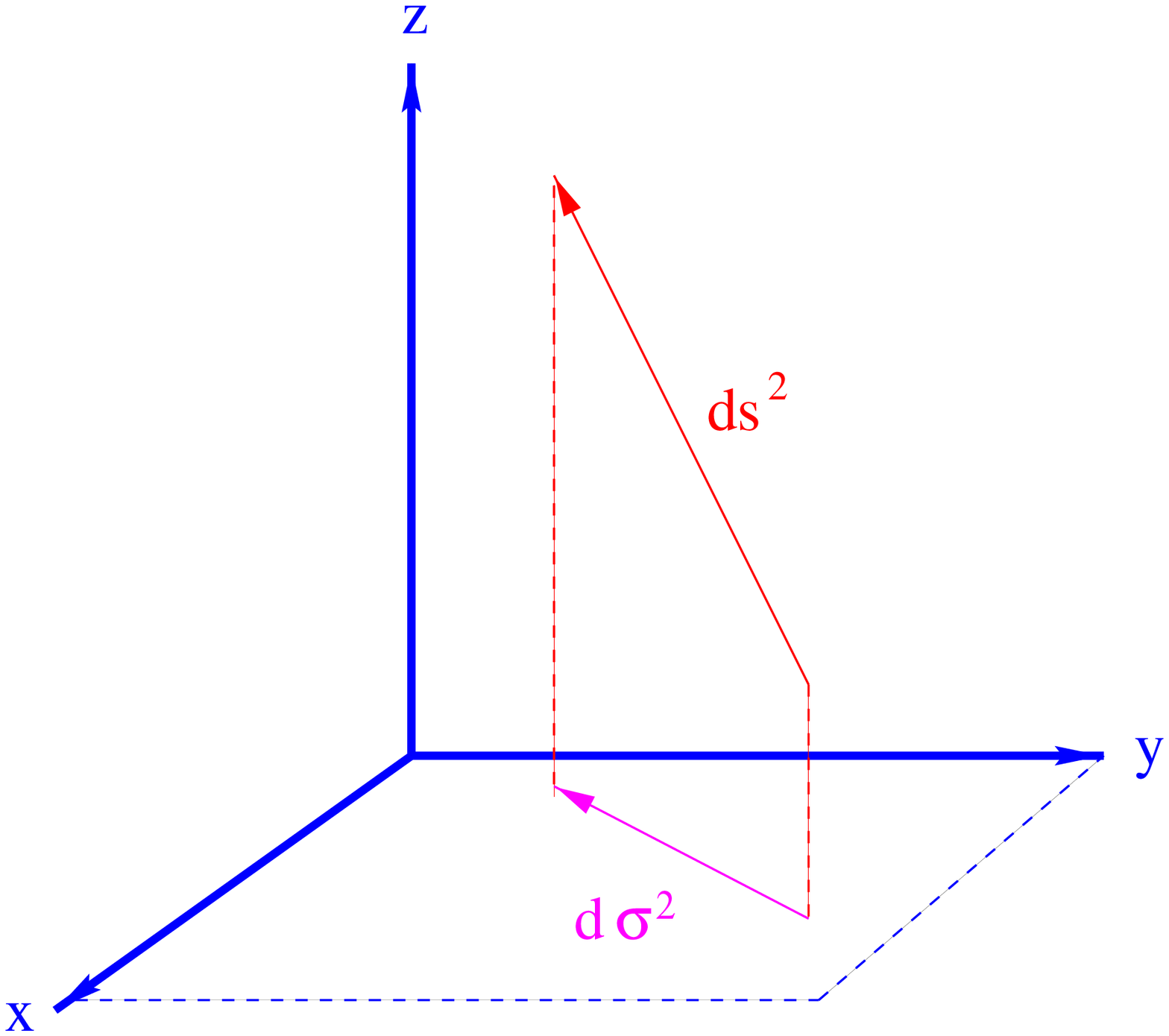}}
\vspace{1cm}
\caption{The minimization of template distance by means of projection onto orthogonal subspaces.}
\label{fig:projectfig}
\end{center}
\end{figure}

From a detection point of view we have seen in Section~\ref{sec:geometry} that the overlap between two templates yields the expression
\begin{equation}
\left<u(\lambda^{\beta})\left|u(\lambda^{\beta}+\Delta\lambda^{\beta})\right>\right. = 1 - \frac{1}{2}\, \xi_{\mu\nu} \Delta\lambda^{\mu}\Delta\lambda^{\nu}.
\label{eq:olap3}
\end{equation}
The left hand side of Eqn~(\ref{eq:olap3}) is maximized if the 2nd term on the right hand side is minimized, i.e. 
\begin{equation}
\Rightarrow \frac{\partial}{\partial\,\Delta\lambda^{\mu}}\left(\xi_{\mu\nu} \Delta\lambda^{\mu}\Delta\lambda^{\nu}\right)  = 0.
\end{equation}
From Eqn~(\ref{eq:propdistderiv}), this process is then equivalent to the minimization of template distance shown above.  We can thus equate the projection of the metric from one subspace onto another as a maximization over the parameters defining each subspace.

\section{Template Number For a One-Step GW Search.}
\vspace{5mm}
\begin{table}[h]
\begin{tabular}{|l|c c | c c|}
\,\,\,\,\,\,\,Detector\,\,\,\,\,\,\, & \,\,\,\,\,\,\,$T_{2}$\,\,\,\,\,\,\, & \,\,\,\,\,\,\,$P_{2}$\,\,\,\,\,\,\, & \,\,\,\,\,\,\,$T_{2.5}$\,\,\,\,\,\,\, & \,\,\,\,\,\,\,$P_{2.5}$\,\,\,\,\,\,\,\\ \hline
\,\,\,\,\,\,\,LIGO\,\,\,\,\,\,\, & \,\,\,\,\,\,\,\,\,$9.77\times10^{3}$\,\,\,\,\,\,\,\,\, & \,\,\,\,\,\,\,\,\,$1.07\times10^{4}$\,\,\,\,\,\, & \,\,\,\,\,\,\,\,$5.64\times10^{3}$\,\,\,\,\,\,\,\, & \,\,\,\,\,\,\,\,$1.47\times10^{4}$\,\,\,\,\,   \\
\,\,\,\,\,\,\,GEO \,\,\,\,\,\,\, & $1.57\times10^{4}$ & $1.71\times10^{4}$ & $9.13\times10^{3}$ & $2.1\times10^{4}$ \\
\,\,\,\,\,\,\,VIRGO\,\,\,\,\,\,\, & $1.26\times10^{5}$ & $1.38\times10^{5}$ & $7.67\times10^{4}$ & $1.62\times10^{5}$ \\
\,\,\,\,\,\,\,TAMA\,\,\,\,\,\,\, & $1.34\times10^{3}$ & $2.26\times10^{3}$ & $7.92\times10^{2}$ & $2.27\times10^{3}$\\
\end{tabular}
\vspace{10mm}
\caption{The number of templates needed to search for inspiralling binaries using T and P-approximant waveforms at 2 and 2.5-PN approximations assuming a minimal match of 0.97.  The minimum individual mass is $1\,M_{\odot}$ and the total mass range is $m\in\left[2, m^{max}\right]M_{\odot}$ }
\label{tab:tappnuma}
\end{table}

\begin{table}[h]
\begin{tabular}{|l|c c | c c|}
\,\,\,\,\,\,\,Detector\,\,\,\,\,\,\, & \,\,\,\,\,\,\,$T_{2}$\,\,\,\,\,\,\, & \,\,\,\,\,\,\,$P_{2}$\,\,\,\,\,\,\, & \,\,\,\,\,\,\,$T_{2.5}$\,\,\,\,\,\,\, & \,\,\,\,\,\,\,$P_{2.5}$\,\,\,\,\,\,\,\\ \hline
\,\,\,\,\,\,\,LIGO\,\,\,\,\,\,\,\,\,\,\,\,\,\,\, & \,\,\,\,\,\,\,\,\,\,\,\,\,\,\,\,\,\,\,\,$51$\,\,\,\,\,\,\,\,\,\,\,\,\,\,\,\,\,\,\,\, & \,\,\,\,\,\,\,\,\,\,\,\,\,\,\,\,\,\,\,\,$75$\,\,\,\,\,\,\,\,\,\,\,\,\,\,\,\,\,\, & \,\,\,\,\,\,\,\,\,\,\,\,\,\,\,\,\,$26$\,\,\,\,\,\,\,\,\,\,\,\,\,\,\,\,\, & \,\,\,\,\,\,\,\,$92$\,\,\,\,\,   \\
\,\,\,\,\,\,\,GEO \,\,\,\,\,\,\, & $94$ & $134$ & $49$ & $161$ \\
\,\,\,\,\,\,\,VIRGO\,\,\,\,\,\,\, & $808$ & $1.02\times10^{3}$ & $439$ & $1.15\times10^{3}$ \\
\,\,\,\,\,\,\,TAMA\,\,\,\,\,\,\, & $16$ & $31$ & $7$ & $42$\\
\end{tabular}
\vspace{10mm}
\caption{The number of templates needed to search for inspiralling binaries using T and P-approximant waveforms at 2 and 2.5-PN approximations assuming a minimal match of 0.97.  The minimum individual mass is $5\,M_{\odot}$ and the total mass range is $m\in\left[10, m^{max}\right]M_{\odot}$ }
\label{tab:tappnumc}
\end{table}

\section{CPU Requirements For a One-Step GW Search.}

\begin{table}[h]
\begin{tabular}{|l|c c | c c|}
\,\,\,\,\,\,\,Detector\,\,\,\,\,\,\, & \,\,\,\,\,\,\,$T_{2}$\,\,\,\,\,\,\, & \,\,\,\,\,\,\,$P_{2}$\,\,\,\,\,\,\, & \,\,\,\,\,\,\,$T_{2.5}$\,\,\,\,\,\,\, & \,\,\,\,\,\,\,$P_{2.5}$\,\,\,\,\,\,\,\\ \hline
\,\,\,\,\,\,\,LIGO\,\,\,\,\,\,\, & \,\,\,\,\,\,\,\,\,$2.12\times10^{9}$\,\,\,\,\,\,\,\,\, & \,\,\,\,\,\,\,\,\,$2.32\times10^{9}$\,\,\,\,\,\, & \,\,\,\,\,\,\,\,$1.22\times10^{9}$\,\,\,\,\,\,\,\, & \,\,\,\,\,\,\,\,$3.19\times10^{9}$\,\,\,\,\,   \\
\,\,\,\,\,\,\,GEO \,\,\,\,\,\,\, & $3.4\times10^{9}$ & $3.71\times10^{9}$ & $1.98\times10^{9}$ & $4.55\times10^{9}$ \\
\,\,\,\,\,\,\,VIRGO\,\,\,\,\,\,\, & $3.08\times10^{10}$ & $3.37\times10^{10}$ & $1.87\times10^{10}$ & $3.96\times10^{10}$ \\
\,\,\,\,\,\,\,TAMA\,\,\,\,\,\,\, & $2.57\times10^{8}$ & $4.33\times10^{8}$ & $1.396\times10^{8}$ & $4.38\times10^{8}$\\
\end{tabular}
\vspace{10mm}
\caption{The CPU requirements (in flops) needed to search for inspiralling binaries using T and P-approximant waveforms at 2 and 2.5-PN approximations assuming a minimal match of 0.97.  The minimum individual mass is $1\,M_{\odot}$ and the total mass range is $m\in\left[2, m^{max}\right]M_{\odot}$ }
\label{tab:cputappa}
\end{table}

\begin{table}[h]
\begin{tabular}{|l|c c | c c|}
\,\,\,\,\,\,\,Detector\,\,\,\,\,\,\, & \,\,\,\,\,\,\,$T_{2}$\,\,\,\,\,\,\, & \,\,\,\,\,\,\,$P_{2}$\,\,\,\,\,\,\, & \,\,\,\,\,\,\,$T_{2.5}$\,\,\,\,\,\,\, & \,\,\,\,\,\,\,$P_{2.5}$\,\,\,\,\,\,\,\\ \hline
\,\,\,\,\,\,\,LIGO\,\,\,\,\,\,\, & \,\,\,\,\,\,\,\,\,$1.31\times10^{7}$\,\,\,\,\,\,\,\,\, & \,\,\,\,\,\,\,\,\,$1.93\times10^{7}$\,\,\,\,\,\, & \,\,\,\,\,\,\,\,$6.68\times10^{6}$\,\,\,\,\,\,\,\, & \,\,\,\,\,\,\,\,$2.36\times10^{7}$\,\,\,\,\,   \\
\,\,\,\,\,\,\,GEO \,\,\,\,\,\,\, & $2.42\times10^{7}$ & $3.44\times10^{7}$ & $1.26\times10^{7}$ & $4.41\times10^{7}$ \\
\,\,\,\,\,\,\,VIRGO\,\,\,\,\,\,\, & $2.3\times10^{8}$ & $2.91\times10^{8}$ & $1.25\times10^{8}$ & $3.27\times10^{8}$ \\
\,\,\,\,\,\,\,TAMA\,\,\,\,\,\,\, & $3.71\times10^{6}$ & $7.18\times10^{6}$ & $1.62\times10^{6}$ & $9.73\times10^{6}$\\
\end{tabular}
\vspace{10mm}
\caption{The CPU requirements (in flops) needed to search for inspiralling binaries using T and P-approximant waveforms at 2 and 2.5-PN approximations assuming a minimal match of 0.97. The minimum individual mass is $5\,M_{\odot}$ and the total mass range is $m\in\left[10, m^{max}\right]M_{\odot}$  }
\label{tab:cputappc}
\end{table}

\end{document}